\documentclass[12pt]{article}
\usepackage{amsmath}%
\usepackage{amsfonts}%
\usepackage{amssymb}%
\usepackage{graphicx}
\begin{document}
\title{Biham-Middleton-Levine Traffic Model With Origin-Destination Trips}
\author{Najem Moussa \thanks{e-mail: najemmoussa@yahoo.fr}
\\\textit{ESNM, D\'{e}pt. de Physique, FST, }
\\\textit{B.P. 509, Boutalamine, Errachidia, Morocco}}
 \maketitle
\begin{abstract}
We extended the Biham-Middleton-Levine model to incorporate the
origin and destination effect of drivers trips on the traffic in
cities. The destination sites are randomly chosen from some
origin-destination distances probability distribution "ODDPD". We
use three different distributions: exponential, uniform and
power-law. We consider two variants of the model. In conserved
particles model (Model A), drivers continue their travelling even
if they reached their destinations. In non-conserved particles
model (Model B), a driver which reaches its destination disappears
with rate $\beta$. It is found that the traffic dynamics in model
A and the evacuation processes in model B are greatly influenced
by the ODDPD. On one hand, we found that we can adjust the ODDPD
to enhance the road capacity of the city and to minimize the
arrival times of drivers in particles conserved system and to
optimize the evacuation time of drivers in non-conserved case. On
the other hand, we find that, independently on the ODDPD, the
evacuation time $T$ of drivers diverges in the form of a power law
$T \propto \beta^{-\nu}$, with $\nu=1$.
\newline\ Pacs numbers: 45.70.Vn, 02.50.Ey, 05.40.-a
\newline\ \textit{Keywords:} Traffic; Cellular Automata;
Freely Moving; Jamming; Arrival Times, Evacuation Times
\end{abstract}
\section{Introduction}
Ever since O. Biham, A.A. Middleton and D. Levine elaborated their
traffic model (BML) \cite{Biham}, the two-dimensional cellular
automata (CA) traffic models are used enormously in order to
understand the complex dynamic behavior of the traffic in cities
(see the reviews [2-4]). In CA, time and space are discrete. The
space is represented as a uniform lattice of cells with finite
number of states, subject to a uniform set of rules, which drives
the behavior of the system. These rules compute the state of a
particular cell as a function of its previous state and the state
of the neighboring cells.
\newline\ The BML model is a three-state CA with periodic boundary
conditions where at a given time step each site can be occupied by
a car moving upward, a car moving rightward, or the site is empty.
There are equal number of cars in each direction and the dynamics
is governed by synchronous traffic lights that allow alternatively
vertical or horizontal movements. At their turn each car jumps to
its next place whenever it is empty. Thus, an essential ingredient
of the BML model is the excluded volume at crossings which
determines the sharp transition between two phases. A first order
phase transition from the free phase to the completely jamming
phase occurs as the car density of the system increases. Recently,
many generalizations and extensions of the BML model are
elaborated to take into account several realistic features of
traffic in cities. For example, the effect of asymmetry
distributions of vehicles among the east-bound and north-bound
streets are investigated in Ref. \cite{Nag2}. Cuesta et al.
\cite{Cuesta} introduced the turn capability of urban car movement
in cities and they found a first order phase transition from a
freely moving regime to a jammed state. Tadaki and Kikuchi
\cite{Tada} found two types of jam phases in the BML model, the
self-organized jam at relatively low density and the random jam at
high density. The anisotropy effect of directions of move on the
BML model, in the periodic and open boundaries conditions are
investigated in Ref \cite{Ez}. Chopard et al. \cite{Chop} have
developed a more realistic CA model of city traffic where the
stretches of the streets in between successive crossings appear
explicitly. The rule for implementing the motion of the vehicles
at the crossing is formulated assuming a rotary to be located at
each crossing. The BML model has been extended also to incorporate
the effects of the static hindrances or road blocks (e.g.,
vehicles crashed in traffic accident) \cite{Nag3,Gu}. It was shown
that the dynamical jamming transition occurs at lower density of
cars with increasing delay time of a car passing over the position
of the traffic accident. J. Freund and T. P\"{o}schel \cite{Fre}
studied another generalization of the BML model. Each site of this
generalized model represents a crossing where a finite number of
cars can wait approaching the crossing from each of the four
directions. The mean velocity  of cars, as a function of the
global traffic density, is determined numerically and derived
analytically applying combinatorics and statistical methods. The
spatial extension of the streets between two intersections is
completely neglected in the original model. To incorporate this
feature, D. Chowdhury and A. Schadschneider \cite{chow2} proposed
a new cellular automata model for vehicular traffic in cities by
combining ideas borrowed from the BML model of city traffic and
the Nagel-Schreckenberg model \cite{ns} of highway traffic. It was
found a phase transition from the free-flowing dynamical phase to
the completely "jammed" phase at a vehicle density which depends
on the time periods of the synchronized signals and the separation
between them \cite{Schad,Bro}.
\newline\ In Sec. $2$, we describe our models for traffic of
 cars in cities with origins and destinations. In Sec. $3$, we
present our numerical results where we give the mean velocity
diagrams of the model A. A detailed description of the
distributions of the arrival times of drivers are also presented.
For model B, we present results concerning the evacuation
processes of drivers inside the lattice. Finally, we conclude with
some conclusions in Sec. $4$.
\section{The BML Traffic Model With Origin-Destination Trips}
In the original BML Traffic Model, cars move, whenever the traffic
lights allow it, all along the lanes without any destinations. In
this paper, we extend the model to describe cars movements with
origin-destination trips. Each car is associated with some given
origin-destination sites. The distance $\delta$ between origin and
destination sites is chosen from certain probability distributions
"ODDPD". We shall consider here three types of distributions:
\newline\ 1) The exponential distribution,
\begin{equation}\label{1}
    f^{e}(\delta)=\frac{\mu}{e^{-\mu \delta_{m}}-e^{-\mu \delta_{M}}}e^{-\mu \delta}
\end{equation}
\newline\ Here we focus the case where $\mu =0.1$ and we suppose
that $f(\delta)$ has a support on some interval
$[\delta_{m},\delta_{M}]$. The numerical values of these
parameters will be defined below.
\newline\ 2) The power-law distribution,
\begin{equation}\label{1}
    f^{p}(\delta)=\frac{n+1}{(\delta_{M}-\delta_{m})^{n+1}}(\delta-\delta_{m})^{n}
\end{equation}
\newline\ Here we focus the case of $n=2$ and we suppose
that $f(\delta)$ has a support on the interval
$[\delta_{m},\delta_{M}]$.
\newline\ 3) The uniform distribution,
\begin{equation}\label{1}
    f^{u}(\delta)=1/(\delta_{M}-\delta_{m})
\end{equation}
\newline\ Here $f^{u}(\delta)$ has a support on the interval $[\delta_{m},\delta_{M}]$.
\newline\ The distance $\delta$ between origin and destination sites, used
all along this paper is defined as follows. Let $(x_{o},y_{o})$
and $(x_{d},y_{d})$ be the coordinates of the origin and
destination sites respectively. We define also the variables
$\delta_{x}$ and $\delta_{y}$ as:
\begin{equation}\label{}
    \left\{%
\begin{array}{ll}
    \delta_{x}= x_{d}-x_{0} \qquad \qquad  \texttt{if}\qquad (x_{d}-x_{0})\geq0 \hbox{;} \\
    \delta_{x}= L+(x_{d}-x_{0})\qquad \texttt{if}\qquad(x_{d}-x_{0})<0 \hbox{.} \\
\end{array}%
\right.
\end{equation}
Similar equations can lead to defining $\delta_{y}$ with replacing
$x$ by $y$. The distance $\delta$ is then given by,
\begin{equation}\label{}
    \delta=\delta_{x}+\delta_{y}
\end{equation}
Initially, all the origins sites of drivers are chosen randomly
from the lattice points. According to traffic lights that permit
horizontal motion at even time steps and vertical motion at odd
time steps, cars will move towards their destinations by selecting
some appropriates paths. Suppose that a car is initially moving
rightward (upward). If the origin and destination sites are in the
same lane (column), then the car will move horizontally
(vertically) until it reaches its destination. Otherwise, the car
will move horizontally (vertically) until it reaches the site
belonging to the column (lane) of the destination site where the
car turns and moves upward (rightward) until it reaches its
destination (see figure 1). In this paper, we shall consider two
simplest versions of the model, which we name model A and model B.
In conserved particles model (model A), cars travel from the
origin sites towards the destination sites. A car which reaches
its destination site continues to travel from that site to a new
destination site, chosen according to the ODDPD, and so on. In
non-conserved particles model ( model B), cars which completes its
journey disappears with certain probability $\beta$, or continues
to travel from that site to a new destination site (as in model A)
with probability $1-\beta$.
\section{Simulation experiments and results}
We carry out our computer simulations of the model by considering
a square lattice of size $L$. Initially, we put randomly a number
$N$ of cars into the lattice with half arrows up and half right.
The density of cars is denoted as $\rho=N/L^{2}$. We use a
parallel updates scheme and a periodic boundary conditions. The
average velocity $<v>$ of cars is defined as the rate of moving
cars to the total number of cars allowed to move by the traffic
lights. The velocity of each car can be either 1 or 0. We perform
numerical simulations of the CA model starting with a set of
random initial conditions for the system size $L=40-200$. After a
transient period that depends on the system size, on the random
initial configuration and on the density of cars, the system
reaches its asymptotic state. The duration of each simulation run
is $50,000$ time steps with the first $20,000$ time steps to
initiate the simulation and the latter $30,000$ used to generate
performance statistics. Finally, we define the parameter
$\delta_{m}$ (resp. $\delta_{M}$) of the support interval of the
ODDPD as the minimal (resp. maximal) distance between the origin
and the destination sites, of the travelling paths of drivers. In
our numerical simulations, we set $\delta_{m}=20$ cells and
$\delta_{M}=2*(L-1)$ cells. The first numerical value is chosen
without any specification while the second rises from the nature
of the path selection procedure followed in this paper (see for
example the gray car and its gray destination in figure 1).
\subsection{Conserved particles model (Model A)}
\subsubsection{Velocity diagrams}
In model A, cars travel from origins to destinations. After the
arrival, the destination sites will be considered as new origins,
from where the travelling will continue towards some new chosen
destination sites, and so on. In figure 2, we carried out the
plots of the mean velocity of cars as a function of the density,
for the three ODDPD. This diagram shows the presence of the two
known states, namely the moving phase and the completely jammed
state where all cars are blocked. In the freely moving state,
interaction between cars is weak and the propagation is important
inside the lattice. In contrast, for large density, the
interaction becomes strong and jamming takes place where car
movements become rare. In contrast with the original BML model,
where a sharp transition separates the two phases, our model shows
that the mean velocity goes down to zero gradually with increasing
the density and vanishes above the transition point. On one hand,
the power-law and the uniform ODDPD lead to almost the same curves
of the plot of $\langle v \rangle$. On the other hand, the
exponential ODDPD leads to some differently curve. So, for low
density the mean velocity of cars of the exponential ODDPD is
smaller than that corresponding to power-law (or uniform) and
higher for high density. Now, let us study the transition point
$\rho_{t}$ separating the moving phase and the completely jammed
state where the average velocity vanishes. From figure 2, we see
clearly that $\rho_{t}$ resulting from the exponential ODDPD is
higher than that resulting from the power-law or the uniform ones.
Hence, the capacity of traffic in cities can be improved if the
travelling paths of drivers are likely shorts.
\newline\ To study the finite size effect on the velocity
diagrams of the system, we compute the mean velocity for different
lattice sizes. That is in contrast with the original BML traffic
model, where the transition point decreases when increasing the
system size $L$ ($\rho_{t}\propto L^{-0.14}$) \cite{shi}, the
velocity diagrams in our models are unchanged with respect to
large lattice sizes (Figure 3a,b). The exponential ODDPD exhibits
an interesting dependence of the velocity diagram on the lattice
size $L$ (Fig.3a). At low densities, $<v>$ remains unchanged with
respect to the variation of $L$. However, in the high density
region, one may distinguish some particular density where the mean
velocity does not depend once again on $L$. Below (resp. above)
this density, $<v>$ decreases (resp. increases) with $L$ and
reaches a stationary value at larger sizes ($L>120$). One sees
from figure 3b that a similar behaviour may occurred for the
power-law ODDPD with the only difference is that the dependence on
$L$ is inverted.
\subsubsection{Arrival time distributions}
When dealing with a realistic traffic in cities, it is important to know the time
necessary to travel from a given origin to a given destination, which is called
"arrival time". This time is interesting to the drivers because it determines when
they must leave their house in order to be on time at their work.
\newline\ In figure 4, we give the arrival time probability distribution
of drivers where the chosen ODDPD is exponential. When the cars
density is low, times shorter than $\tau_{1}\sim2*\delta_{m}$ are
strongly suppressed while beyond the probability decays
exponentially and exhibits a single peak at $\tau_{1}$. Since the
traffic light turns periodically and the minimal distance taken by
each driver is supposed equal to $\delta_{m}=20$ cells, the
shortest arrival time will be closely equal to $40$ \emph{time
steps}. In addition, since cars interact very weakly at low
densities, the arrival time of any driver is almost equal to
$2*\delta$ where $\delta$ is the origin-destination distance
chosen by a driver. This justifies the fact that the arrival time
distribution keeps the same shape (exponential) as the ODDPD. For
higher densities, the probability distribution of arrival times
presents a single peak at time approximately equal to $\tau_{1}$;
reflecting the short routes chosen by almost all drivers. Beyond
$\tau_{1}$, the arrival time distribution decreases exponentially.
Consequently, with exponential ODDPD, arrival times of drivers
will be likely shorts even if the density is relatively high. In
another hand, there are no significant changes of the probability
distribution of arrival times with respect to relatively large
lattice sizes and relatively low densities. This result can be
clarified from two points. First, only small distances from
origins to destinations can be drawn from the exponential
distribution of equation (1). Second, all the origins points of
drivers are chosen randomly from the lattice points. Hence,
standing times lost by a driver are usually shorts since the
interaction with others cars is weak. Nevertheless, when the cars
density is close to $\rho_{t}$, cars self-organized into a large
cluster (large jam) and almost all vehicles do not move. Thus,
drivers situated in the interior of this large jam do not move and
rest inside for an infinite time. The only moving cars are
situated at the boundaries of the large jam which can reach their
destinations at finite times.
\newline\ In figure 5, we plot the arrival time probability distribution
of drivers where the ODDPD is power-law. As before, when the cars
density is low, the form of the arrival time distribution is
similar to that of the ODDPD because the system is in free regime.
However, in contrast to the exponential ODDPD, the power-low one
exhibits a strong dependence on the lattice size. In deed, the
distances drawn from the power-law ODDPD are usually large;
leading therefore to an enhancement of the arrival times of
drivers if the lattice size is increased. If the car density
increases, one finds evidently a broad distribution of arrival
times, because cars are moving for quite long times before
reaching their destinations. The higher is $\rho$, the higher is
the arrival time. Yet, the arrival time distribution exhibits two
peaks. The arrival time at the first peak which is located at
$\tau_{2}\sim2*\delta_{M}$ changes little with increasing the
density. Beyond $\tau_{2}$, the distribution spreads to the higher
value of arrival times. On the other hand, times shorter than
$\tau_{2}$ represent the arrival times of cars which terminate
their journey without that they are stopped. However, times longer
than $\tau_{2}$ correspond to cars which stopped for a while
during their routes. In the long times region, the probability
distribution increases, reaches a maximum and then decreases
asymptotically towards zero. This peak is considered as the most
probable arrival time of a driver who travels in congested region
of the lattice. This behaviour which occurs also for the
time-headway in highways \cite{mou1}, rises from the strong
interactions between cars at high densities. Furthermore, long
arrival times should be taken by drivers in large lattice sizes,
because larger distances could be drawn from the probability
distribution of Eq.$(2)$.
\newline\ If the uniform ODDPD is used and the
cars density is low, one sees that the corresponding arrival times
distribution is uniform (Fig. 6). For higher densities, we find
that the arrival times distribution behaves as the power-law case,
especially in long arrival times region. The main difference is
that, in the power-law case, the distribution presents two peaks
while, in the uniform case, it presents three peaks. The first
peak is the higher corresponding to the arrival time $\tau_{1}$
while the second peak is shifted a little bit towards the higher
values with respect to $\tau_{2}$. Drivers with short trips can
avoid congested regions and move towards their destinations with a
minimal arrival times. However, those which travel far away,
cannot avoid congested regions and their arrival times will be
larger than the distances of their trips. The third peak gives the
most probable arrival time of such drivers.

\subsection{Non-Conserved particles model (Model B)}
In model B, cars move from origins to destinations. In contrast to
model A, cars which completes its journey disappears with certain
probability $\beta$, or continues to move from that site to a new
destination site with probability $1-\beta$. In model B, the
number of cars present in the lattice, $N(t)$, is not conserved
but decreases with time until becomes zero above certain time $T$,
called hereafter as the evacuation time, i.e., the time it takes
for all cars to leave the lattice. Hereafter, we shall investigate
the effect of the ODDPD on $N(t)$.
\newline\ If $\beta=1$, the number of cars remains constant
up to certain time $\tau_{1}\sim40$ \emph{time steps} which is the
same for the three different ODDPD. This is the time needed to
reach a destination for a driver which have an origin-destination
distance equal to $\delta_{m}=20$ cells. Above $\tau_{1}$, the
number of cars $N(t)$ shows a polynomial or a linear or an
exponential decreases if we used respectively a power-law, or a
uniform or an exponential ODDPD. Moreover, as figure 7a shows, the
evacuation time is much shorter for the exponential distribution
than for the two others distributions. As $\beta$ diminishes,
$N(t)$ decreases slowly with time; leading therefore to an
increasing of the evacuation time. Most interestingly, if $\beta$
becomes very small, $N(t)$ will decrease almost exponentially for
the three different ODDPD. In addition, the evacuation times are
found to be very long for the power-law ODDPD (see Fig. 7b).
\newline\ It is found that, when the time $t$ exeeds $\tau_{1}$, the exponential
ODDPD produces an exponential decrease of the number $N(t)$
whatever the values of the system parameters. Thus, $N(t)$ follows
the equation,
\begin{equation}\label{}
N(t)=N_{0}e^{-\lambda t}
\end{equation}
where $N_{0}$ is the initial number of drivers present in the
lattice. Our results depicted in figures 8a and 8b show that the
coefficient $\lambda$ is independent on both the initial density
$\rho_{0}$ and the lattice size $L$. However, it is found that
$\lambda$ depends strongly on $\beta$ and $\lambda=f(\beta)$ is an
increasing function which shows a strong increase as $\beta$
approaches the maximal value ($\beta=1$). This function is
illustrated in figure 8c.
\newline\ Finally, the dependence of the evacuation time
$T$ on the parameter $\beta$ is depicted in figure 9a for the
three ODDPD at low and high densities. We find that, the
evacuation time diverges as $\beta$ approaches the vanishing
value. From figure 9b we see that the evacuation time $T$ follows
a power law behavior of the form,
\begin{equation}\
    T \propto \beta^{-\nu}
\end{equation}
Except for some minor fluctuations, the dynamic exponent $\nu$
remains unchanged when varying the density or when changing the
type of ODDPD. For examples, if the exponential ODDPD is used, one
finds, $\nu\approx 1.08\pm0 .001$ for $\rho_{0}=0.1$ and
$\nu\approx 1.09\pm 0.007$ for $\rho_{0}=0.3$, while for the
power-law case one finds, $\nu\approx 1.10\pm 0.009$ for
$\rho_{0}=0.1$ and $\nu\approx 1.11\pm 0.006$ for $\rho=0.3$.
Assuming that the parameter $\beta$ is rate of transition for the
dynamics of the model B, and as it was demonstrated in Ref.
\cite{mou2} for a one-dimensional traffic model, the exponent
$\nu$ is expected to be theoretically equal to one.
\section{Conclusions}
In summary, we have extended the BML model to incorporate the
origin and destination effect of drivers trips on the traffic in
cities. The origin-destination distances are drawn from certain
probability distribution that is used for all drivers.
\newline\ In model A, all drivers continue their travelling
even if they reached their destinations. The mean velocity of cars
is greatly influenced by the ODDPD. Yet, the transition density
that separates the moving phase and the completely jammed state,
is higher when we used the exponential ODDPD than if we used the
power-law or the uniform ODDPD. The arrival times of drivers
depend on the density and on the type of ODDPD. At low density,
cars interact very weakly and the arrival time is almost equal to
$2*\delta$ where $\delta$ is the origin-destination distance.
However, at high density, cars strongly interact and jamming phase
occurs. This increases the arrival time of some drivers which
stopped for a while during their journeys. Our results indicate
that we can adjust the ODDPD to enhance the road capacity of the
city and to minimize the arrival time of drivers. The effect of
lattice size $L$ is also studied. Hence, in contrast to the
original BML model, the model A shows a non dependence of the mean
velocity on the lattice size beyond some large limit. However, the
distribution of arrival times becomes broad as we increase the
size $L$, especially for the power-law and the uniform distance
distributions.
\newline\ The other version studied in this paper is the model B
where a driver which reaches its destination disappears with rate
$\beta$. We found that the evacuation processes is greatly
influenced by the ODDPD. For the three ODDPD and with respect to
varying the rate $\beta$, the evacuation time is found to exhibit
a power law behaviour. This evacuation processes is characterized
by a dynamical exponent $\nu$ ($\tau\propto \beta^{-\nu}, \nu=1$).

\newpage\

\newpage\ \textbf{Figures captions}
\begin{quote}
\textbf{Figure 1}. Illustration of traffic in a square lattice (with periodic
boundary conditions) where each car moves from origins to destinations. Cars
(rectangles) located at the origin sites will move towards their destination sites
(circles).
\newline\ \textbf{Figure 2}. Mean velocity diagrams of cars versus
density for the three different ODDPD.
\newline\ \textbf{Figure 3}. Mean velocity diagrams of cars versus
density for different values of lattice size $L$. (a) Exponential
ODDPD; (b) Power-law ODDPD.
\newline\ \textbf{Figure 4}. The arrival time probability distribution
of drivers for the exponential ODDPD.
\newline\ \textbf{Figure 5}. The arrival time probability distribution
of drivers for the power-law ODDPD.
\newline\ \textbf{Figure 6}. The arrival time probability distribution
of drivers for the uniform ODDPD.
\newline\ \textbf{Figure 7}. Time evolution of the number of drivers present in the lattice
for the three ODDPD and for various values of the rate $\beta$;
(a) $\beta=1.0$ and $\beta=0.5$; (b) $\beta=0.1$.
\newline\ \textbf{Figure 8}. Time evolution of the number of drivers present in the lattice
for the exponential ODDPD; (a) for various values of the rate
$\beta$ and density $\rho$; (b) for various lattice size $L$.
\newline\ \textbf{Figure 9}. The dependence of the evacuation time $T$ on the
 parameter $\beta$ for the three ODDPD at low and high densities; (a) Linear plots
(b) Log-Log plots.
\end{quote}

\begin{thebibliography}\
\bibitem{Biham}
{O. Biham, A. A. Middleton, and D. Levine, Phys. Rev. A
\textbf{46} 6124 (1992).}
\bibitem{Hel1}
{D. Helbing, Rev. Mod. Phys. \textbf{73} 1067 (2001).}
\bibitem{chow}
 {D. Chowdhury, L. Santen and A. Schadschneider, Phys. Rep.
\textbf{329} 199 (2000).}
\bibitem{nag}
{T. Nagatani, Rep. Prog. Phys. \textbf{65} 1331 (2002).}
\bibitem{Nag2}
{T. Nagatani, J. Phys. Soc. Jap. \textbf{62}, 2656(1993).}
\bibitem{Cuesta}
{J.A. Cuesta, F.C. Martines, J.M. Molera and A. Sanchez, Phys.
Rev. E \textbf{48}, R4175 (1993).}
\bibitem{Tada}
{S. Tadaki and M. Kikuchi, Phys. Rev. E \textbf{50}, 4564 (1994).}
\bibitem{Ez}
{A. Benyoussef, H. Chakib, and H. Ez-Zahraouy, Phys. Rev. E
\textbf{68}, 026129 (2003).}
\bibitem{Chop}
{B. Chopard, P.O. Luthi and P.A. Queloz, J. Phys. A \textbf{29},
2325 (1996).}
\bibitem{Nag3}
{T. Nagatani, J.Phys.A 26, L1015 (1993); J. Phys. Soc. Jap. 62,
1085 (1993).}
\bibitem{Gu}
{G.Q. Gu, K.H. Chung and P.M. Hui, Physica A \textbf{217}, 339
(1995).}
\bibitem{Fre}
{J. Freund and T. P\"{o}schel, Physica A \textbf{219}, 95 (1995); J. Stat. Phys.,
\textbf{86}, 421 (1996).}
\bibitem{chow2}
{D. Chowdhury and A. Schadschneider, Phys. Rev. E \textbf{59},
R1311 (1999).}
\bibitem{ns}
{K. Nagel and M. Schreckenberg, J. Phys. (France) I, \textbf{2},
2221 (1992).}
\bibitem{Schad}
{A. Schadschneider, D. Chowdhury, E. Brockfeld, K. Klauck, L.
Santen, and J. Zittartz, Traffic and Granular Flow '99 Springer,
New York (2000).}
\bibitem{Bro}
{E. Brockfeld, R. Barlovic, A. Schadschneider and M.
Schreckenberg, Phys. Rev. E \textbf{64} 056132 (2001).}
\bibitem{shi}
{Y. Shi, adap-org/9509003.}
\bibitem{mou1}
{N. Moussa, Int. J. Mod. Phys. C \textbf{15}, 29 (2004).}
\bibitem{mou2}
{N. Moussa, Phys. Rev. E \textbf{71}, 026124 (2005).}
\end{thebibliography}
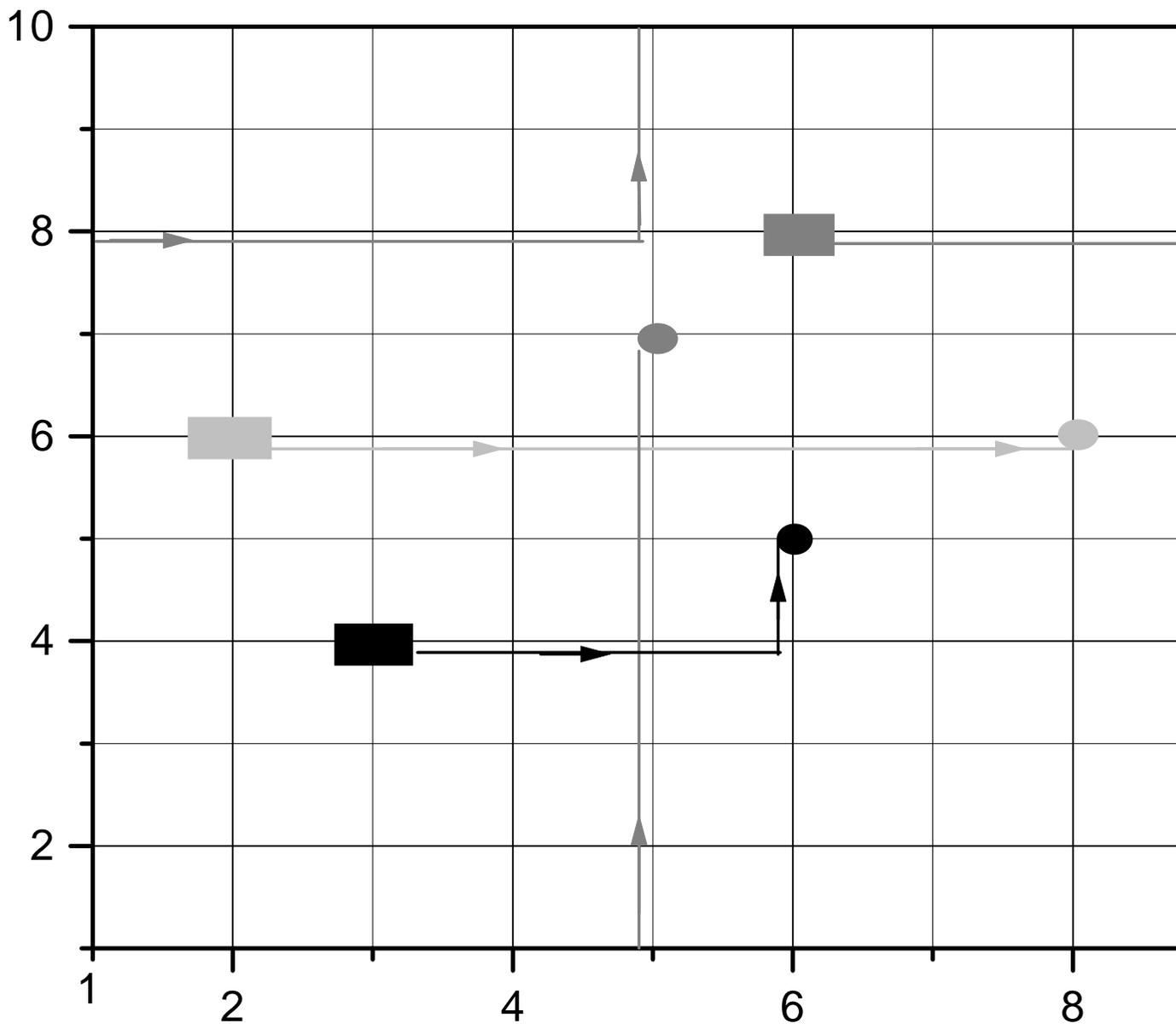
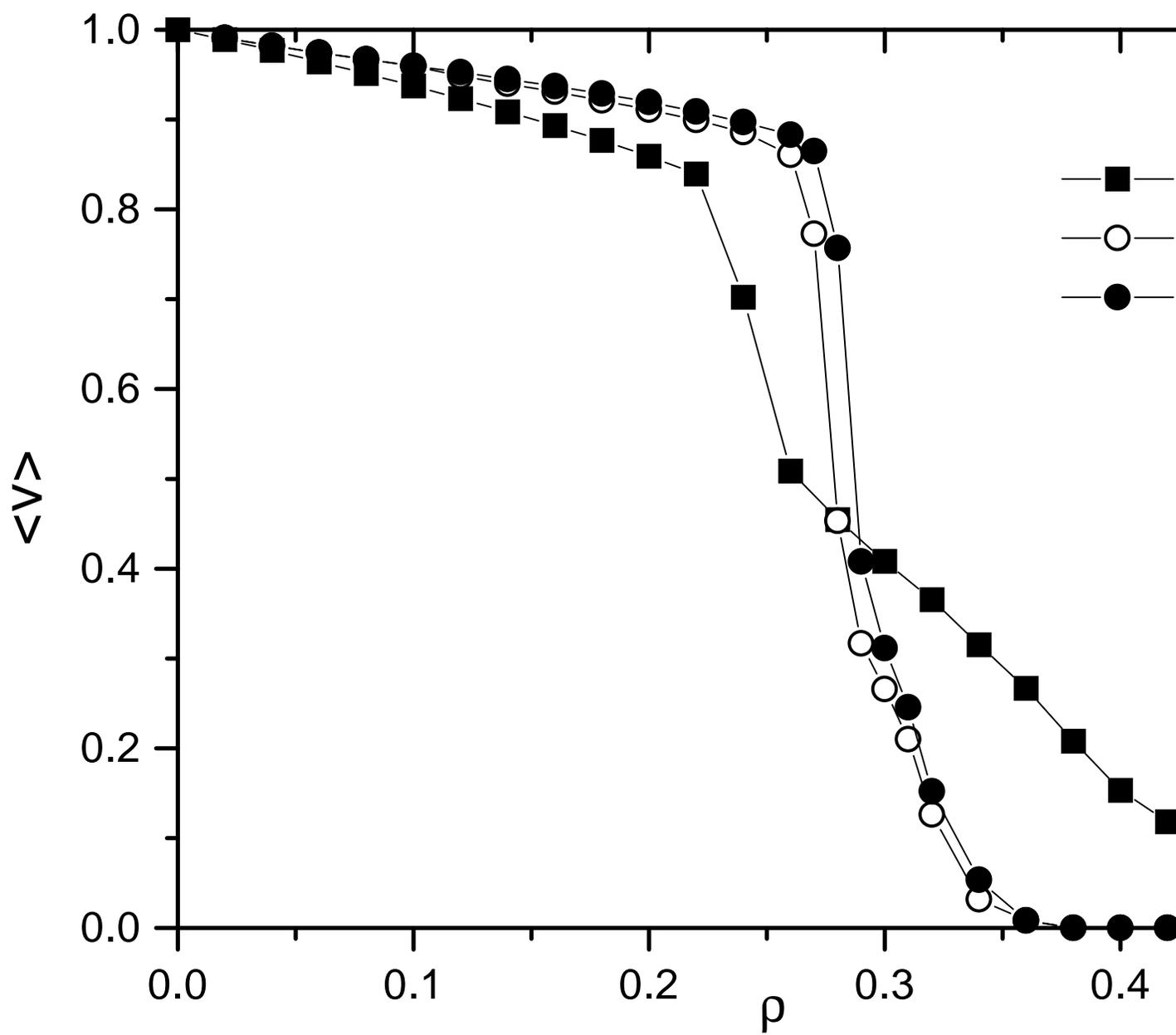
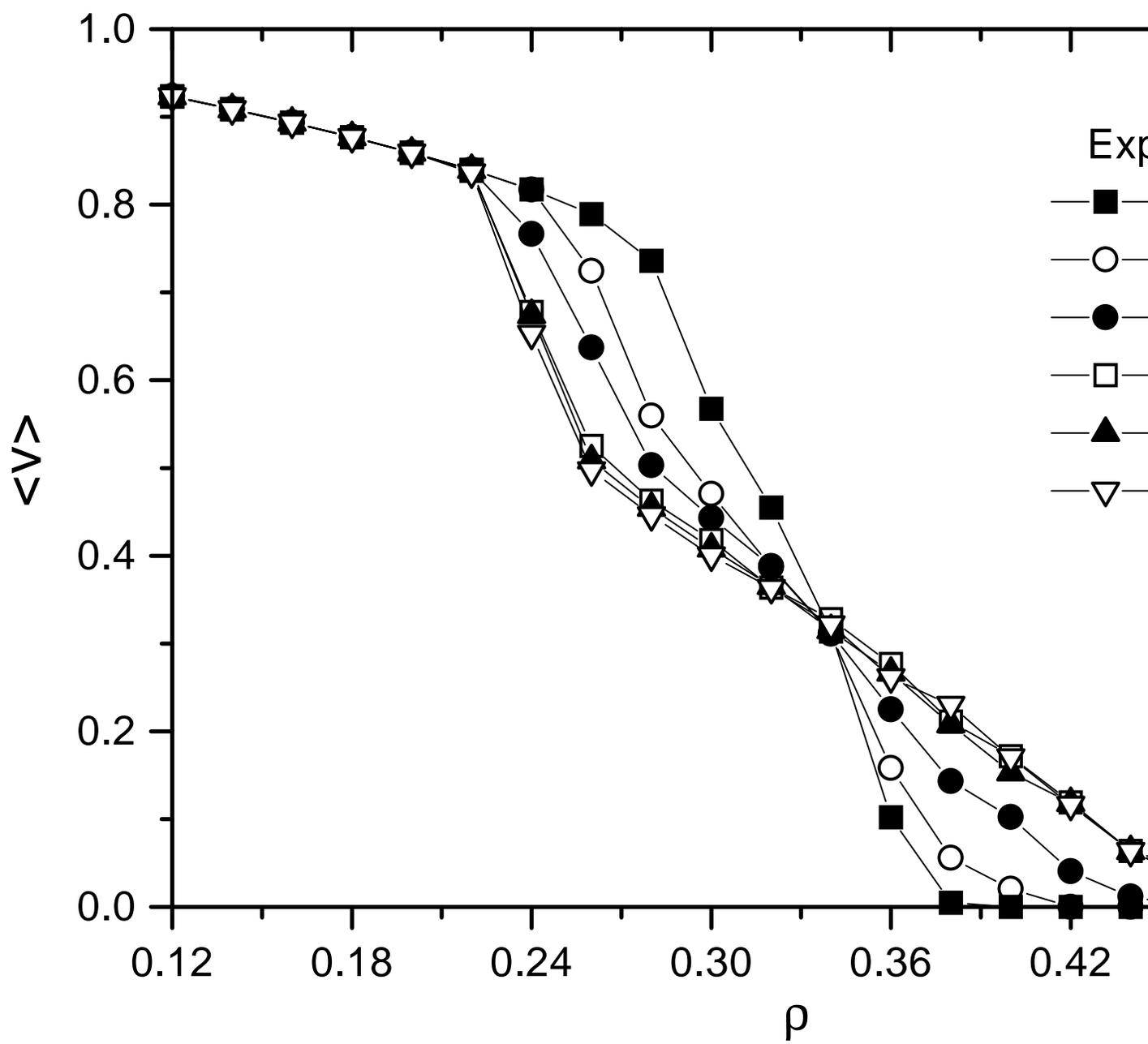
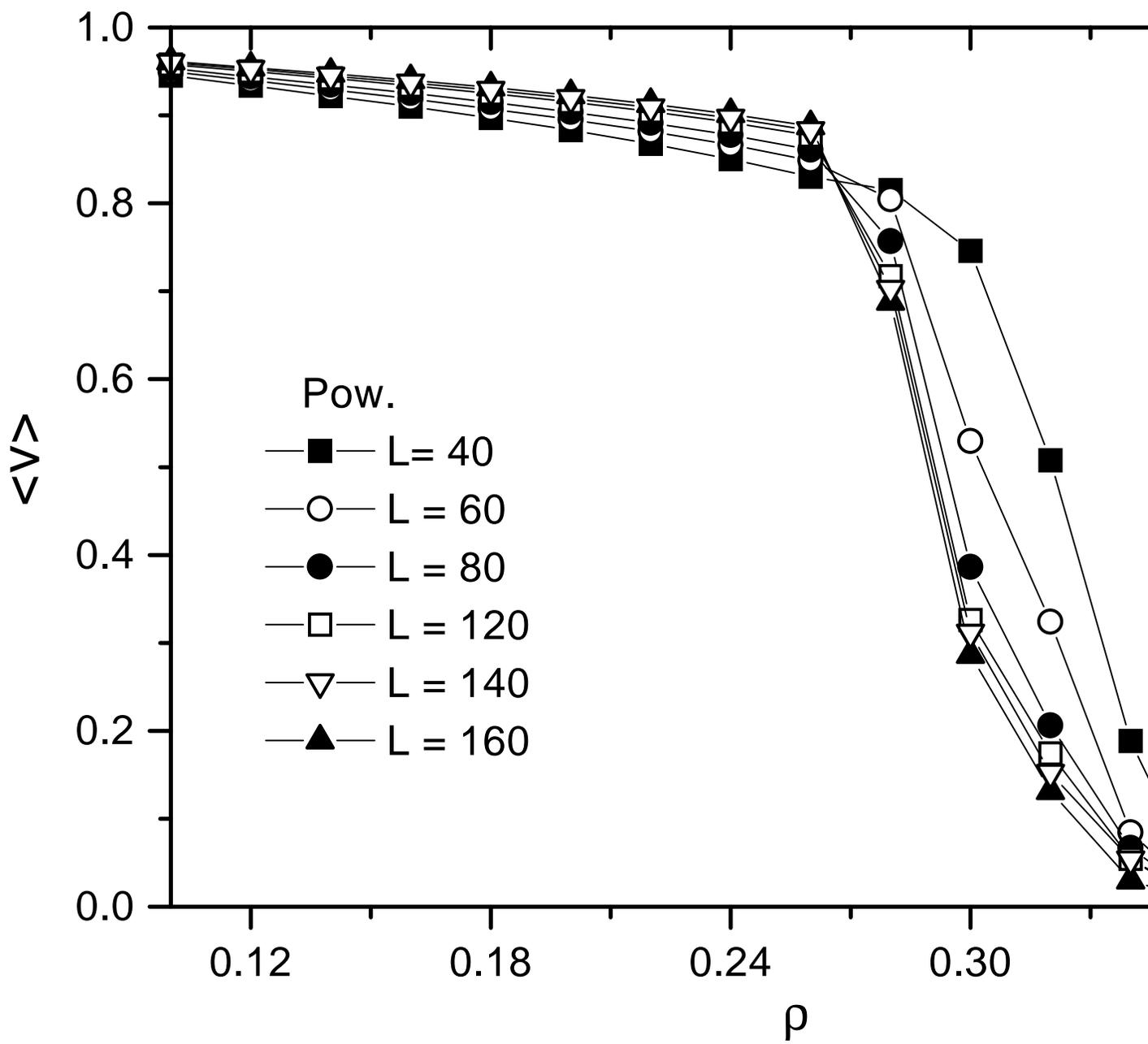
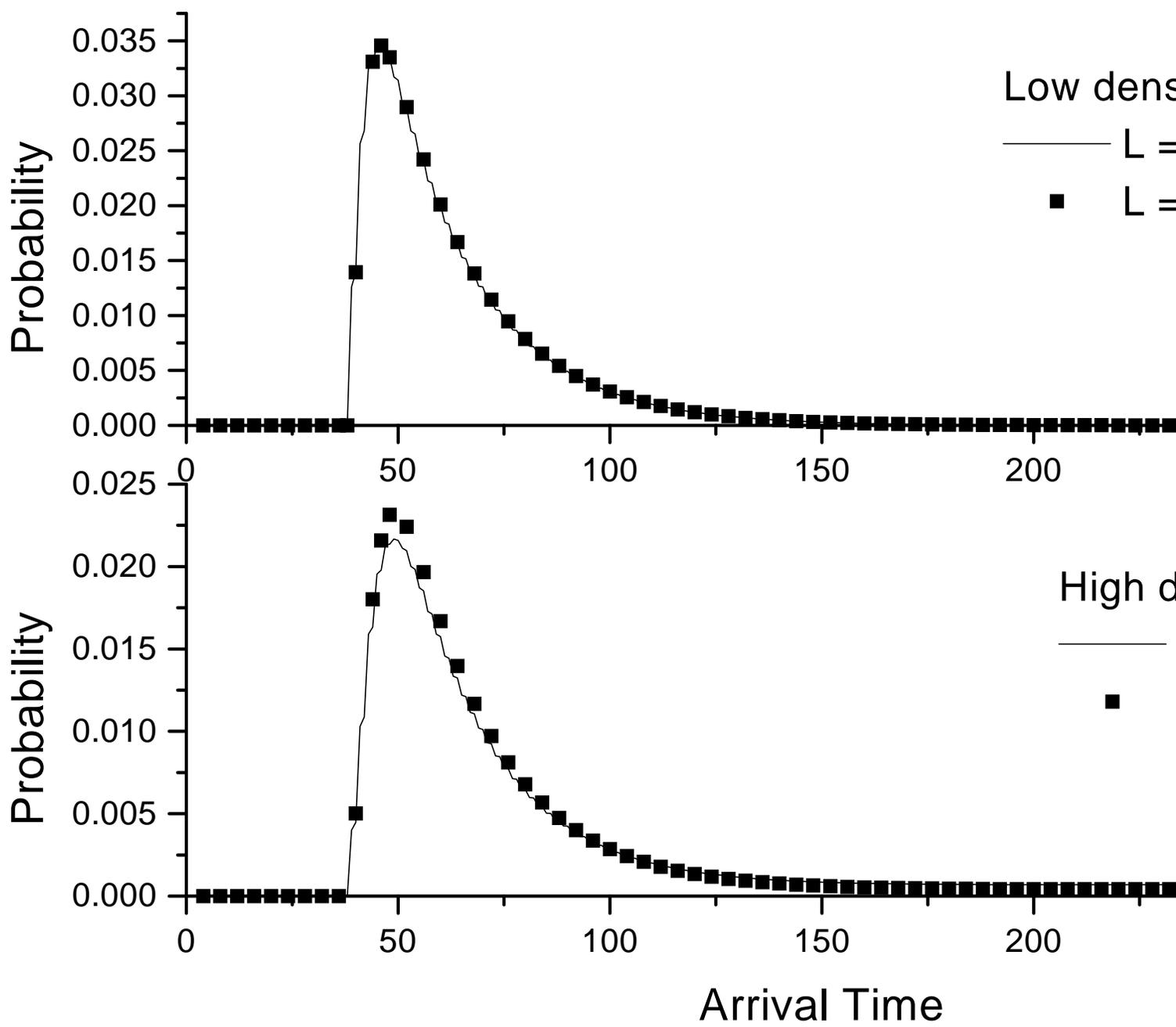
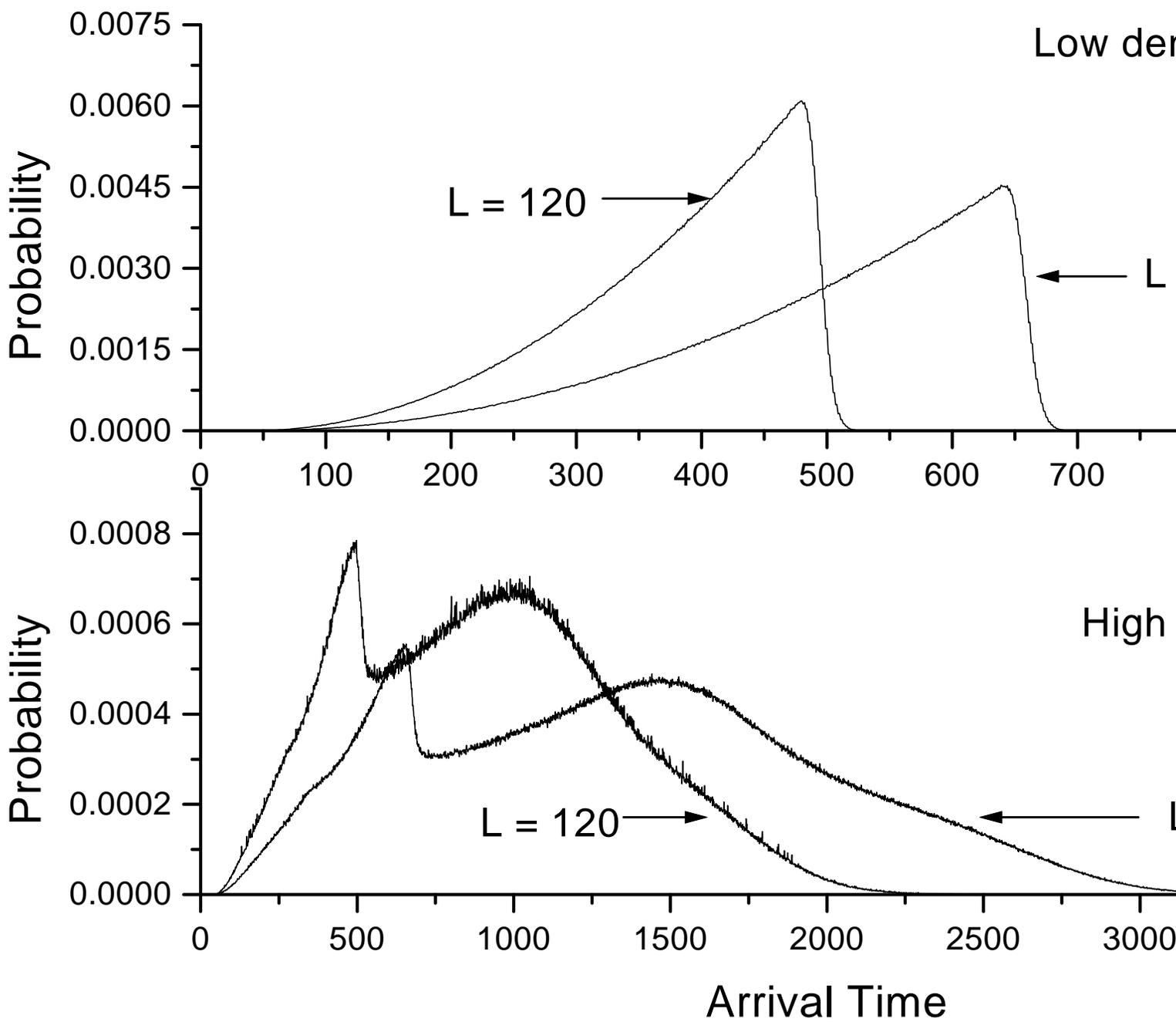
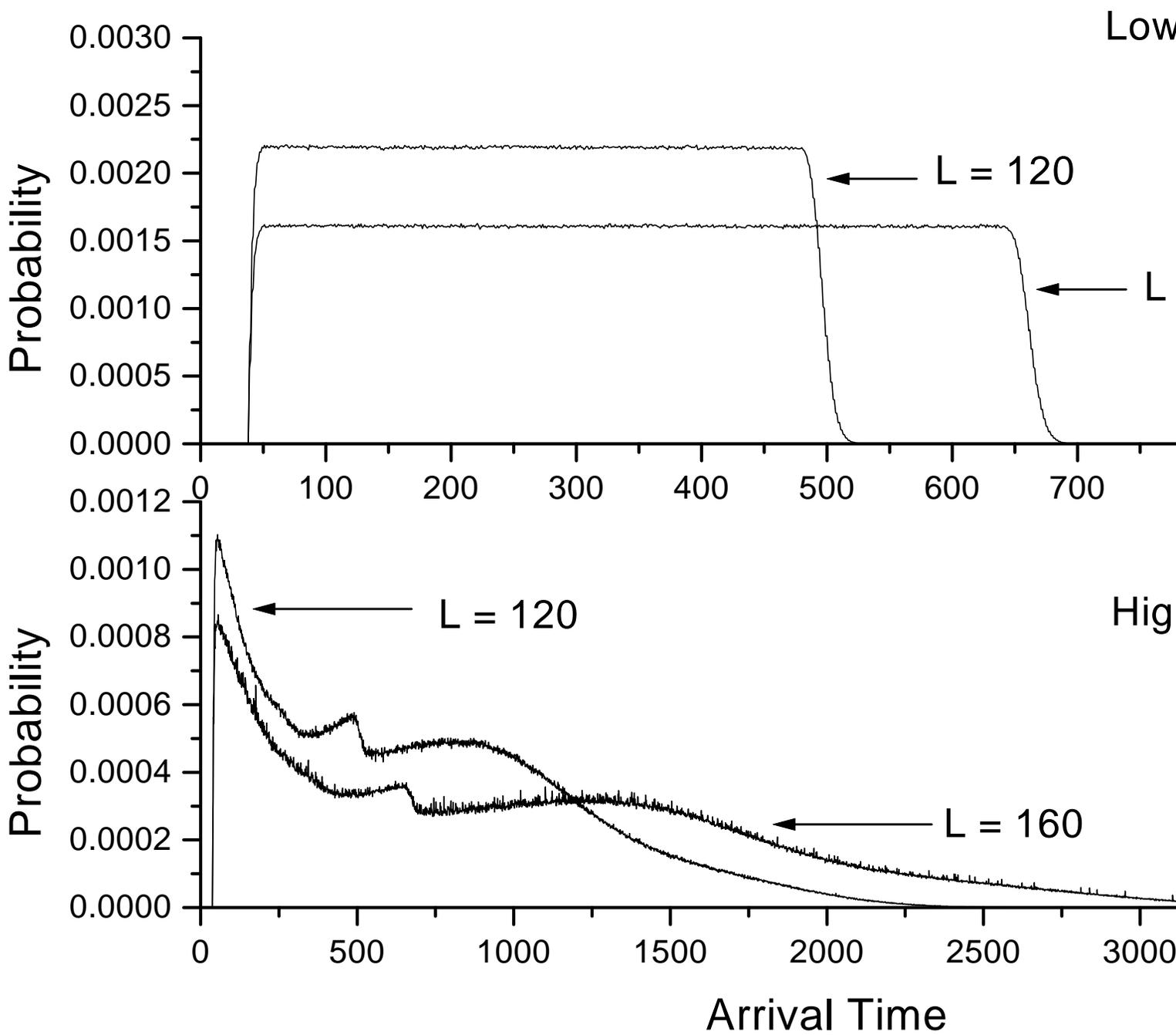
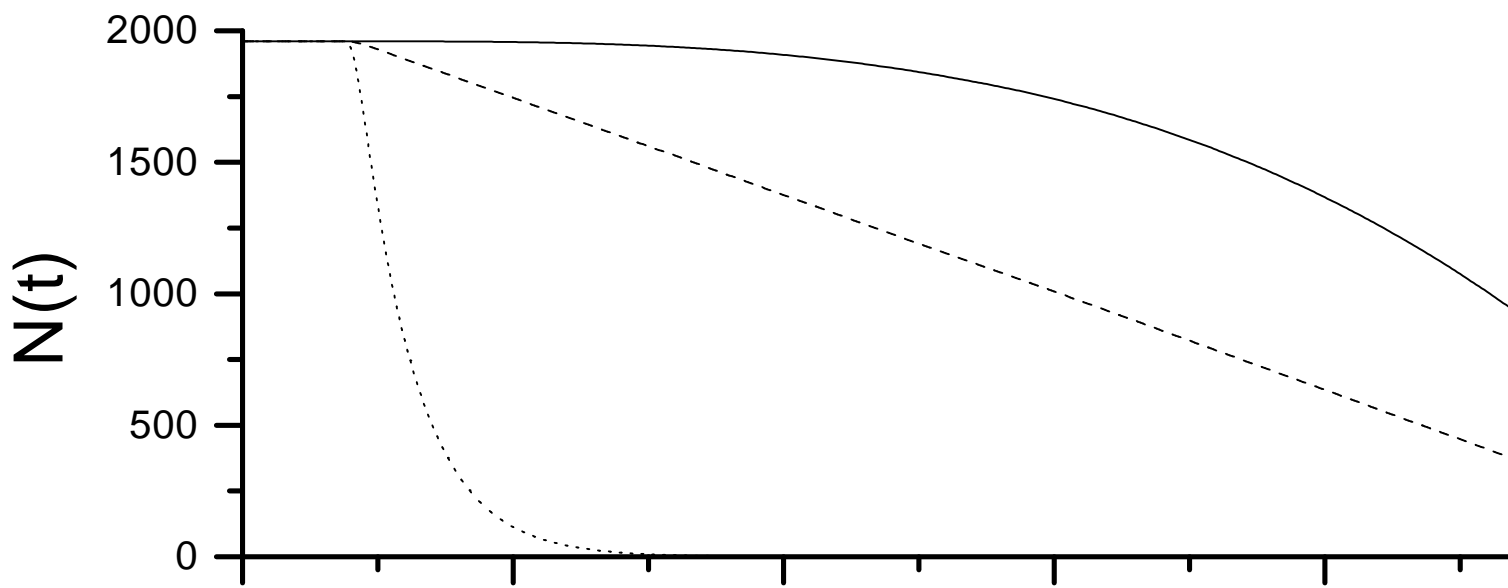
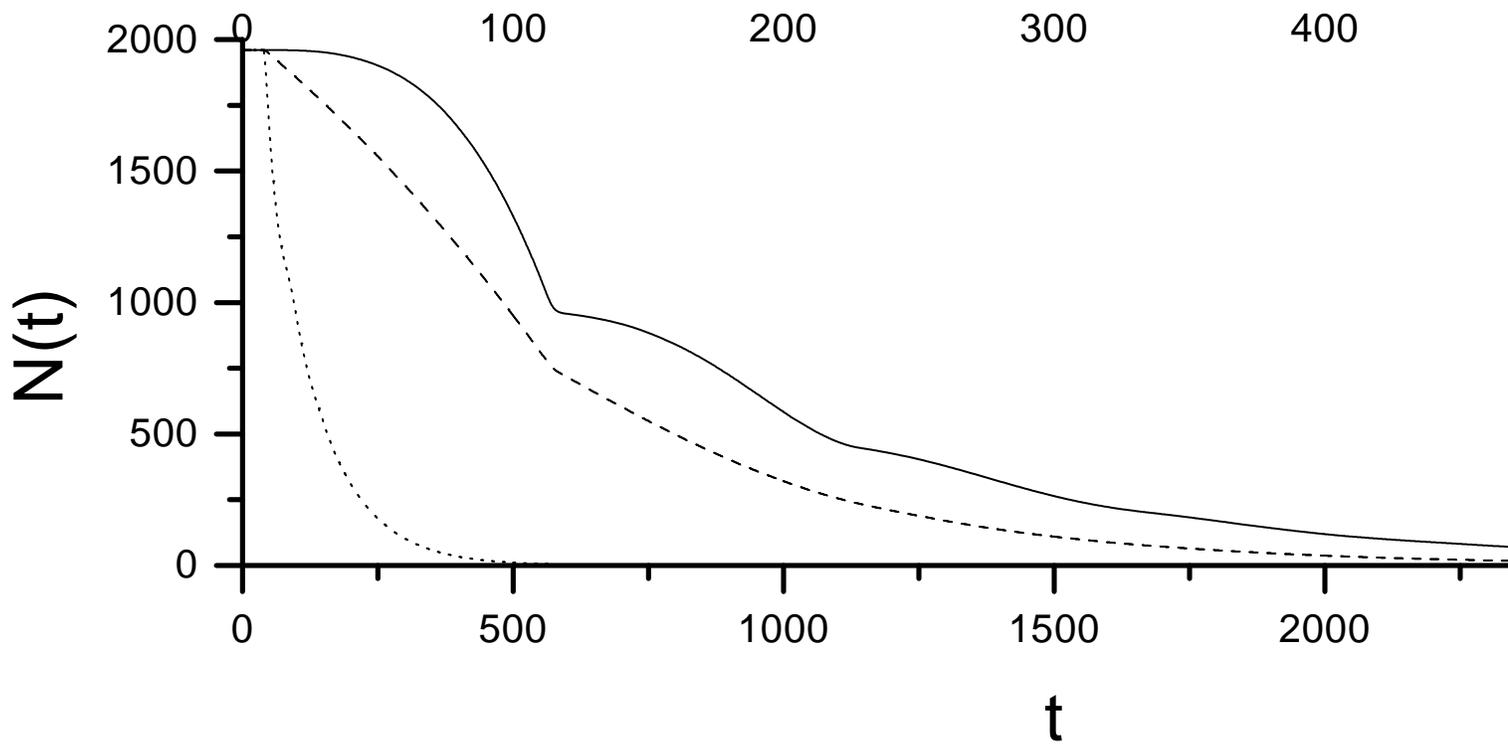
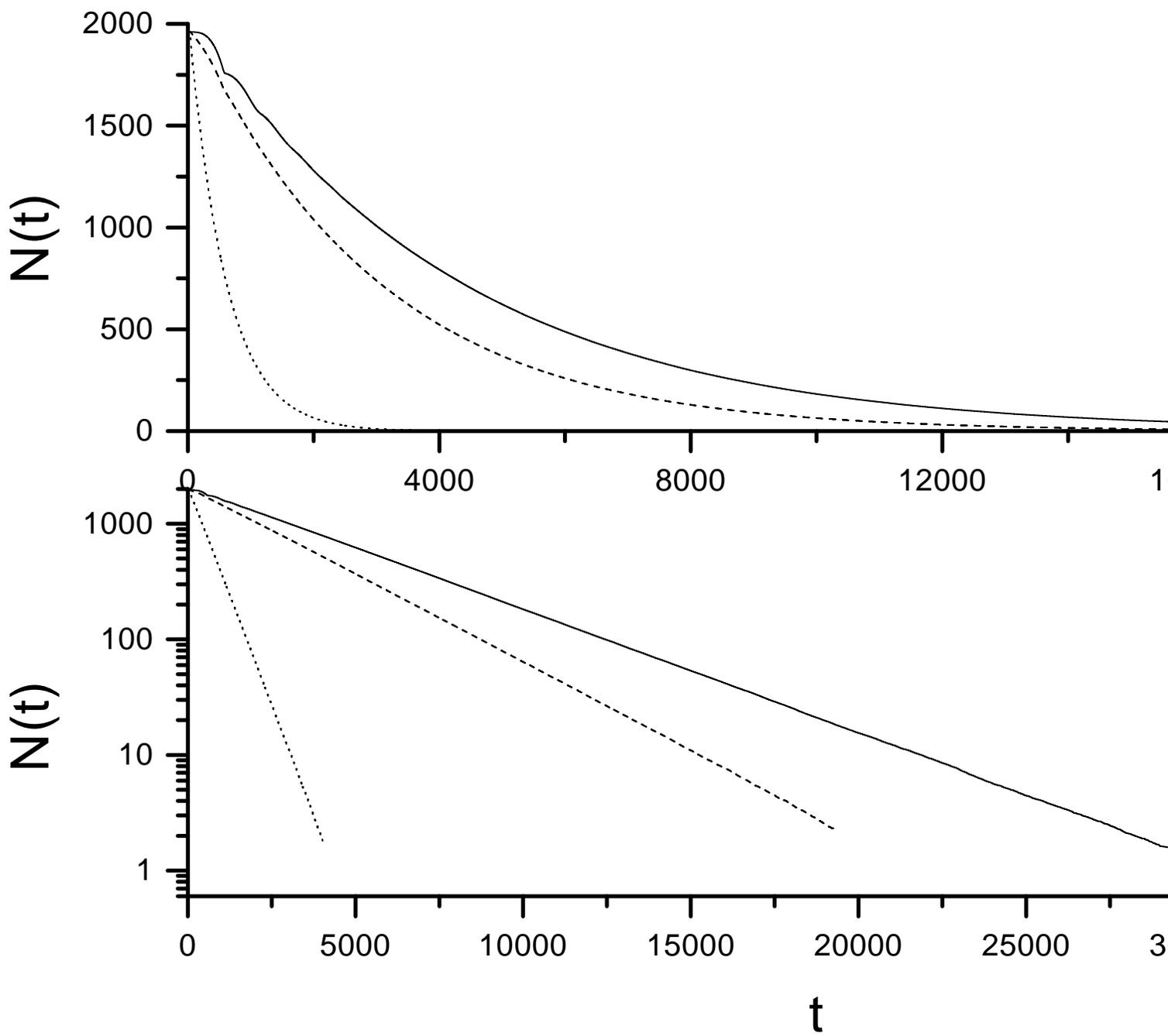
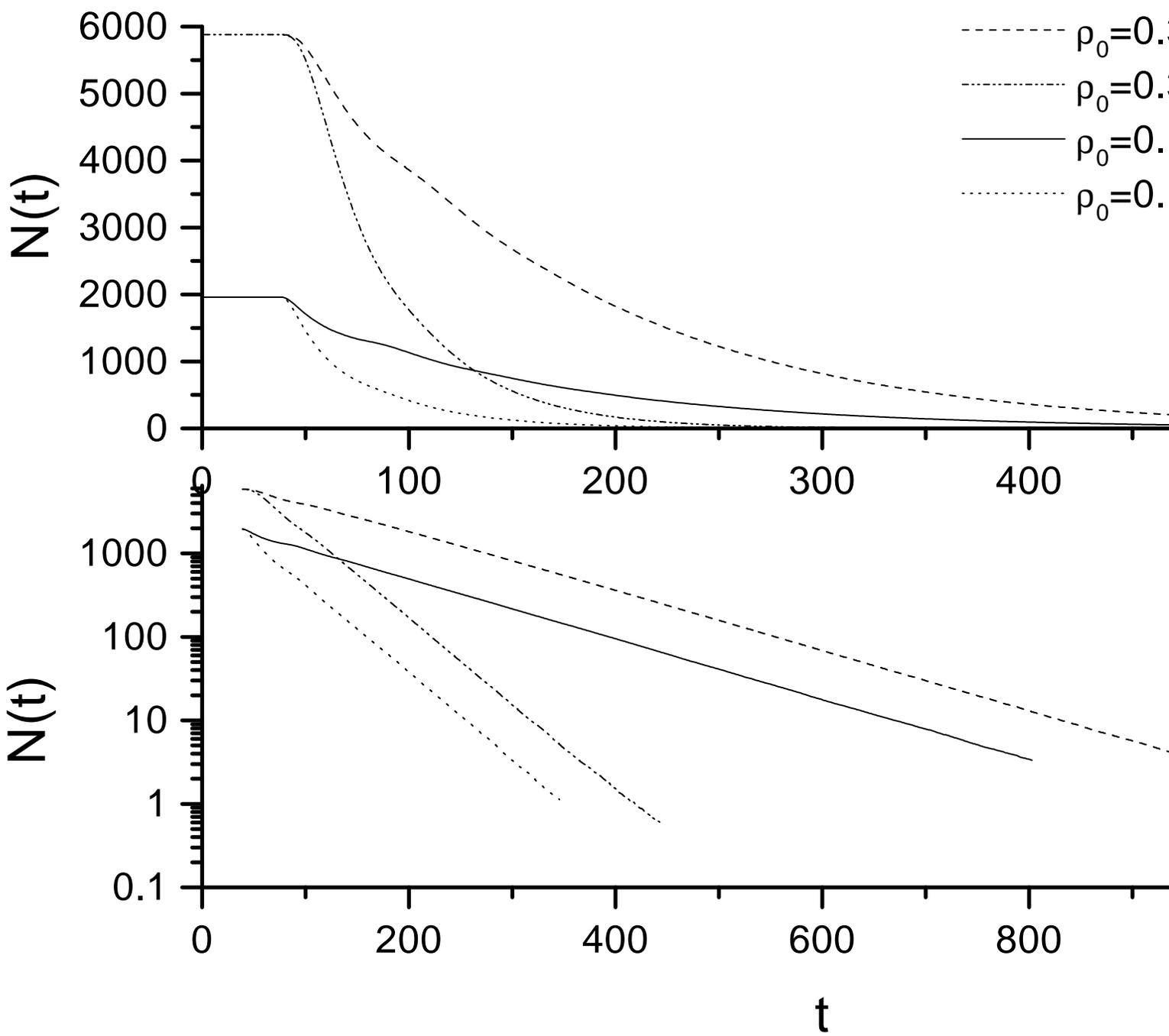
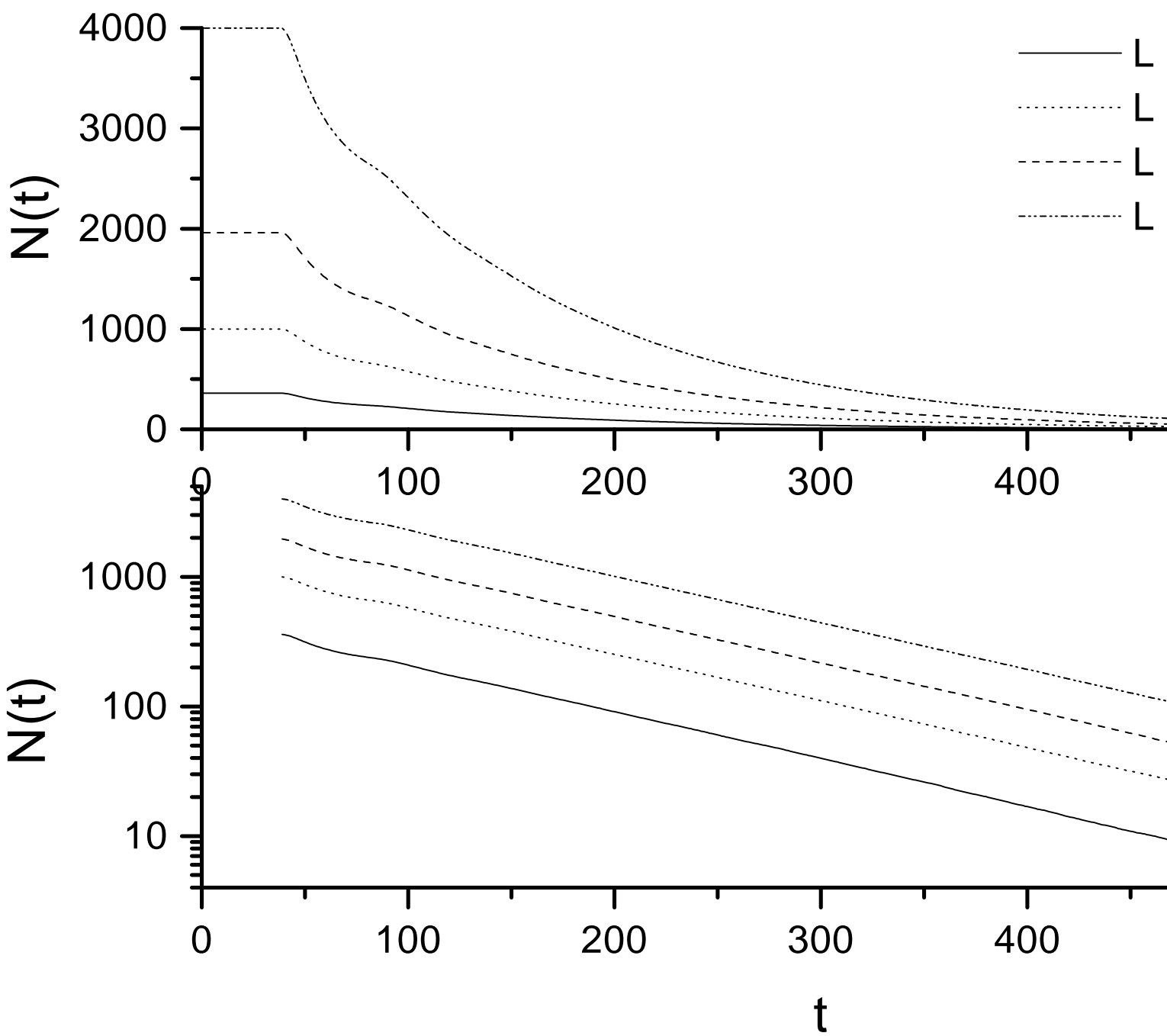
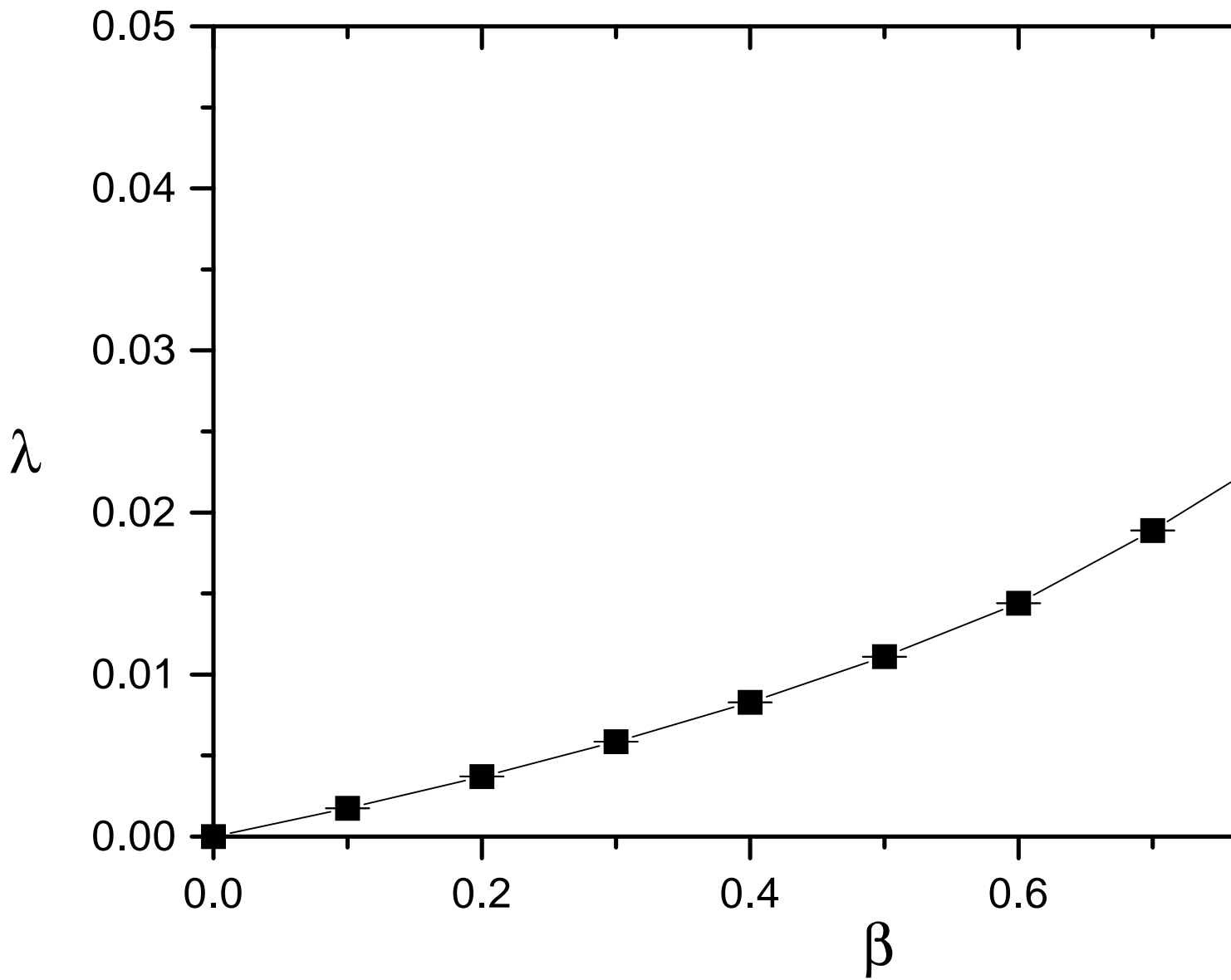
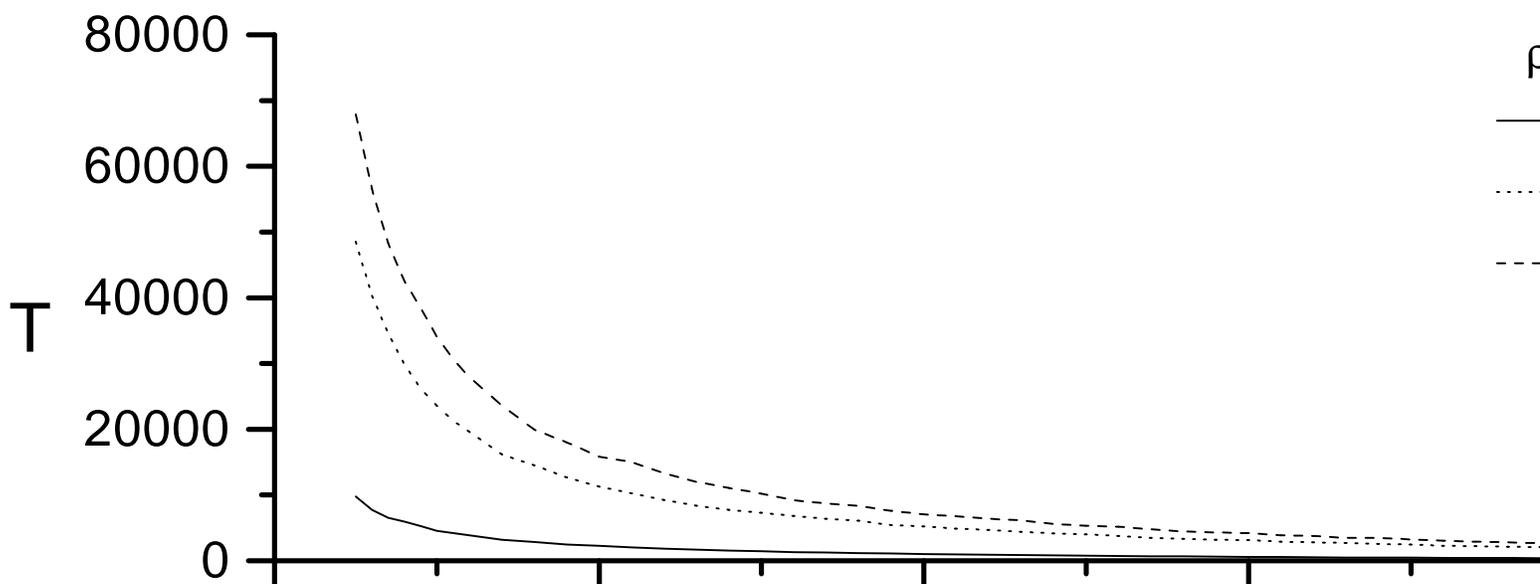
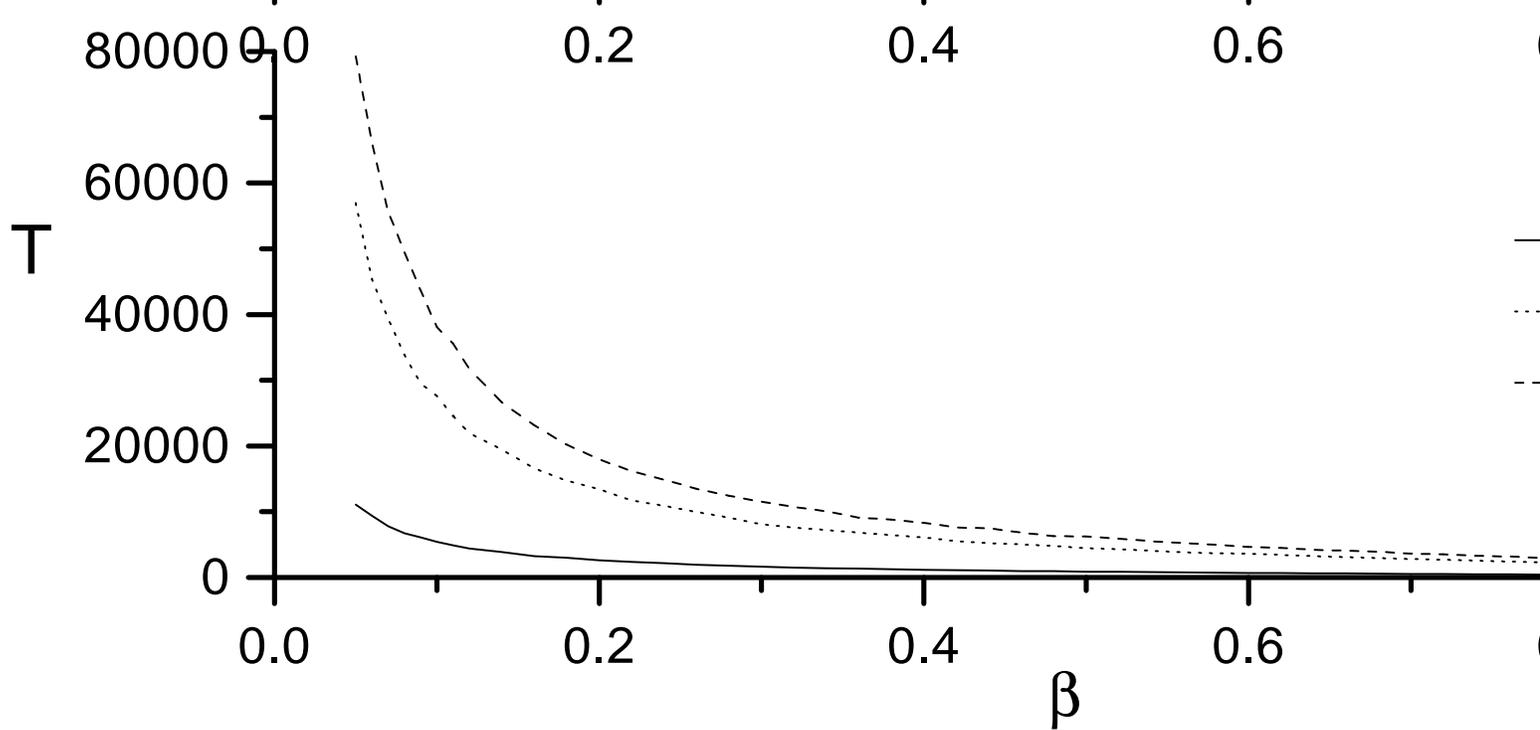
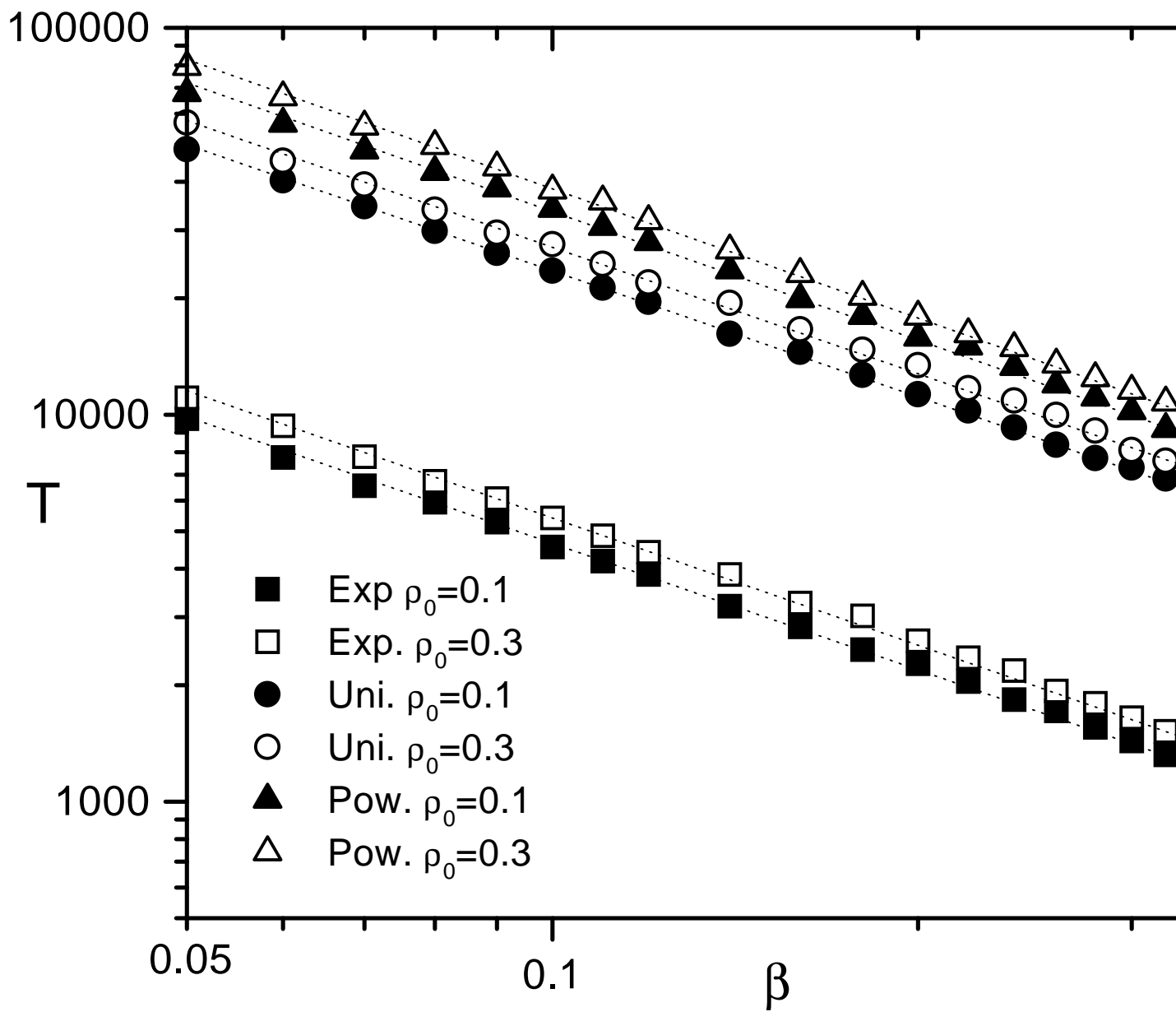
\end{document}